\PassOptionsToPackage{table,xcdraw}{xcolor}
\documentclass[sigconf]{acmart}

\copyrightyear{2022}
\acmYear{2022}
\setcopyright{acmcopyright}\acmConference[MSR '22]{19th International Conference on Mining Software Repositories}{May 23--24, 2022}{Pittsburgh, PA, USA}
\acmBooktitle{19th International Conference on Mining Software Repositories (MSR '22), May 23--24, 2022, Pittsburgh, PA, USA}
\acmPrice{15.00}
\acmDOI{10.1145/3524842.3527975}
\acmISBN{978-1-4503-9303-4/22/05}

\usepackage{graphicx}
\usepackage{caption}
\usepackage[ruled,linesnumbered]{algorithm2e}
\usepackage{booktabs}
\usepackage{subfigure}
\usepackage{url}
\usepackage{listings}

\usepackage{titlesec}
\titlespacing\section{10pt}{4pt plus 1pt minus 4pt}{6pt plus 0pt minus 4pt}
\titlespacing\subsection{0pt}{6pt plus 0pt minus 4pt}{6pt plus 0pt minus 4pt}
\titlespacing\subsubsection{0pt}{6pt plus 0pt minus 4pt}{6pt plus 0pt minus 4pt}

\usepackage{awesomebox} 
\usepackage{fontawesome5}
\definecolor{findingBarColour}{gray}{0.4}
\definecolor{findingIconColour}{gray}{0.3}

\usepackage{tcolorbox}

\tcbuselibrary{skins}
\newtcolorbox{rqbox}[1]
{
  before skip=1em, 
  after skip=1em, 
  colframe = black!60,
  colback  = black!10,
  coltitle = black!90,  
  title    = \textbf{#1},
  hbox boxed title,
  enhanced,
  attach boxed title to top center={yshift=-3mm,yshifttext=-1mm},
  boxed title style={size=small, colback=black!30}
} 

\usepackage{fancybox}
\newenvironment{hassanbox}%
{\begin{center}\vspace{0mm}\noindent\begin{Sbox}\begin{minipage}{0.97\columnwidth}}%
{\end{minipage}\end{Sbox}\fbox{\TheSbox}\end{center}\vspace{0mm}}

\begin{document}

\title{An Empirical Study on Maintainable Method Size in Java}


\author{Shaiful Alam Chowdhury}
\affiliation{%
  \institution{University of British Columbia}
  \city{Vancouver}
  \state{BC}
  \country{Canada}
}
\email{shaifulc@cs.ubc.ca}

\author{Gias Uddin}
\affiliation{%
  \institution{University of Calgary}
  \city{Calgary}
  \state{AB}
  \country{Canada}
}
\email{gias.uddin@ucalgary.ca}
\author{Reid Holmes}
\affiliation{%
  \institution{University of British Columbia}
  \city{Vancouver}
  \state{BC}
  \country{Canada}
}
\email{rtholmes@cs.ubc.ca}
\renewcommand{\shortauthors}{Chowdhury et al.}

\begin{abstract}
Code metrics have been widely used to estimate software maintenance effort.
Metrics have generally been used to guide developer effort to reduce or avoid future maintenance burdens. 
Size is the simplest and most widely deployed metric.
The size metric is pervasive because size correlates with many other common metrics (e.g., McCabe complexity, readability, etc.).
Given the ease of computing a method's size, and the ubiquity of these metrics in industrial settings, it is surprising that no systematic study has been performed to provide developers with meaningful method size guidelines with respect to future maintenance effort.
In this paper we examine the evolution of $\sim$785K Java methods and show that developers should strive to keep their Java methods under 24 lines in length.
Additionally, we show that decomposing larger methods to smaller methods also decreases overall maintenance efforts.
Taken together, these findings provide empirical guidelines to help developers design their systems in a way that can reduce future maintenance.

\end{abstract}




\maketitle

\section{Introduction}

Software maintenance has long been identified as challenge for software engineers~\cite{Kafura:1987} and
maintenance costs often exceed initial  development costs~\cite{mchain:2016}. 
Consequently, both researchers and practitioners are keen to model future maintenance effort from the current state of a software project---to facilitate risk planning and to assist project optimization~\cite{Shin:2011,Zhou:2010,bigbangs:2019,Cruz:2019,Kondo:2020,McClure:1978,TOSUN:2010}.  
These maintenance models rely on various metrics to provide the data necessary to predict future maintenance challenges.

Two of the most widely adopted maintenance indicators are change-proneness~\cite{Romano:2011,Gil:2017} and bug-proneness~\cite{Zimmermann:2007,Islam:2020,PASCARELLA:2020,Menzies:2007,TOSUN:2010}. 
These metrics suggest that
code components (e.g., methods, classes, or modules) that are bug- or change-prone are generally expensive to maintain.
Early identification and optimization of such components could thus help reduce future maintenance effort. 
Unfortunately, change- and bug-proneness of code components are not generally available until after one or more system releases~\cite{Spadini:2018}.

To address this shortcoming, 
researchers have investigated static code metrics to learn how these metrics alias with change- and bug-proneness  (e.g.,~\cite{Kafura:1987,Zhou:2010,Gil:2017,PASCARELLA:2020}). 
These metrics include McCabe~\cite{McCabe:1976}, McClure~\cite{McClure:1978}, nested block depth~\cite{Johnson:2019}, C\&K~\cite{Chidamber:1994}, and size, just to name a few. 
The utility of these metrics, however, has long been debated
(e.g.,~\cite{Shepperd:1988,Gil:2016,Gil:2017}). 
While some research has shown that these static metrics correlate with bug- and change-proneness 
(e.g.,\cite{Bandi:2003,Antinyan:2014,Landman:2014,Spadini:2018, Johnson:2019}), other research has criticized these findings (e.g.,~\cite{Scalabrino:2017,Gil:2017,Shepperd:1988}). 
One underlying concern with these metrics is how they alias with the size of the component being measured.
It has been shown that 
most code metrics are highly influenced by their correlation with size~\cite{Emam:2001,TSE:2013,Gil:2017}. 
This suggests that component size, which is easy for software engineers to both measure and reason about, could be the primary maintenance indicator~\cite{Gil:2017}.
Ultimately, developers want to be able to design their systems in a way that reduces future maintenance effort.
Unfortunately, there is no evidence-based guidelines that developers can follow to optimize the size of their components.
In this paper we focus on method level granularity to measure size (e.g., in contrast to class-, or module-level granularities), because method level granularity is desired by both developers and researchers~\cite{grund:2021, PASCARELLA:2020}. 
There is not currently broad consensus for how long a method should be; for example, consider \emph{`What is the ideal length of a method for you?'}~\cite{SO-method-length}. 
The suggestions vary greatly: between 5-15 lines, 50 lines, and 100-200 lines. 
We propose that having evidence-based guidance to help developers understand the relationship between method size and software maintenance would help them better structure their systems to reduce future maintenance effort. 
Therefore, this paper describes an empirical study to investigate optimal method sizes with respect to software maintenance.

To do this, we examined the evolution of $\sim$785K Java methods from 49 open-source software projects to investigate the relationship between method length and change- and bug-proneness. 
Based on our study, we make the following four primary contributions to the field of software maintenance:

\noindent \textbf{Maintenance effort correlates with method length (RQ1).}
Previous work has shown that both file-length~\cite{Gil:2017} and module-length~\cite{Gill:1991} negatively impact maintenance; we extend these findings to method-length.
We also investigate the best method length measure to use for maintenance prediction.





\noindent \textbf{Methods should be 24 SLOC or fewer (RQ2).} 
Suggesting shorter methods are better is insufficient guidance for practical application as there are challenges to having arbitrarily small methods.
We show that methods that are 24 or fewer SLOC are less maintenance-prone than longer methods.
We also show that in practice this threshold is widely achievable.




\noindent \textbf{Decomposed large methods undergo less maintenance (RQ3).}
Complex problems often require complex solutions, which may exceed the 24 SLOC threshold. 
A natural question is whether these complex solutions are inherently maintenance-prone. 
We show that in most cases, larger methods are more maintenance-prone than methods that have been decomposed into shorter helper methods.

\noindent \textbf{Controlling for size also controls other code quality metrics (RQ4).}
We show that keeping method size within 24 SLOC also improves the testability, readability, and maintainability of source methods.



Taken together, this paper provides an evidence-based analysis of the impact on method length on software maintenance that developers can use to reduce future maintenance effort as they write and evolve their systems.
We further show that these length guidelines are both achievable in practice and suggest that refactoring complex methods into smaller sub-methods further improves system maintainability.
To reproduce our results, we publicly share the dataset of $\sim$785K methods with their change and bug information.\footnote{https://github.com/shaifulcse/MaintainableSLOC-data}

\section{Related Work}
Maintenance effort can be estimated through external factors such as correctness and performance of the underlying software system. Unfortunately, these factors are hard to collect and are often unavailable from the beginning of software development life cycle (SDLC)~\cite{Gil:2017}. Internal metrics, also known as code metrics, are easy to collect and are available throughout the SDLC. Therefore, it has been a holy grail to the research community to understand software maintenance using different code metrics~\cite{McCabe:1976, McClure:1978, Chidamber:1994, Lake:1994, PASCARELLA:2020, Trautsch:2020}. Consequently, a significant numbers of code metrics have been proposed and studied over the last forty years~\cite{Lenarduzzi:2017}. 

McCabe~\cite{McCabe:1976}, or cyclomatic complexity, is a measure of the number of independent paths in a source code component. The assumption is that a source code component with high McCabe score would be hard to maintain; it would be more change- and bug-prone~\cite{Curtis:1979, Zhou:2010, Baggen:2012, Tiwari:2014,  Antinyan:2014, Landman:2014, Antinyan:2015, Pantiuchina:2018, PASCARELLA:2020}. 
Another popular code metric is the Halstead metric, which is mainly based on the number of operators and operands in a code component~\cite{Kafura:1987, Curtis:1979, Hindle:2008, Posnett:2011, Antinyan:2017}. 
Abreu \textit{et al.}~\cite{Abreu:1994} proposed the coupling factor metric which is based on the client-supplier relationships among different classes. 

Chidamber and Kemerer have proposed the popular suite of six OO metrics, known as the C\&K metric~\cite{Chidamber:1994}. This suite includes \textit{Depth of Inheritance Tree}, \textit{Number of Children}, and \textit{Coupling between Object Classes}. Readability as a code metric was developed by Buse \textit{et al.}~\cite{Buse:2010}, who modelled different code features, such as number of comments, and number of blank lines, to a readability score. 
This model was later improved by Posnett \textit{et al.}~\cite{Posnett:2011} and Scalabrino \textit{et al.}~\cite{Scalabrino:2016}. 
Mo \textit{et al}.~\cite{Mo:2016} developed the decoupling level metric to measure how hard it is to modify a class without affecting other classes in a software project. 
Code metrics are often used to define antipatterns (code smells)---i.e., poor design choices that make a software less maintainable. Studies have found that classes (and modules) with antipatterns are generally more change- and bug-prone than classes with no, or few, antipatterns~\cite{Romano:2012, Khomh:2012, Hall:2014}. 

Unfortunately, the effectiveness of code metrics for understanding future maintenance has been debated extensively. 
Although many studies have been positive about the utility of reasoning about maintenance with code metrics~\cite{Johnson:2019, Landman:2014,Spadini:2018,Bandi:2003,Antinyan:2014}, some studies found them ineffective~\cite{Shepperd:1988,Scalabrino:2017, Gil:2017}. According to these studies, size is the only code metric that can estimate maintenance to some extent~\cite{Emam:2001,Gil:2017,TSE:2013, Herraiz:2007}, and all the other code metrics are only as effective as their correlation strength with size~\cite{Gil:2017}. A code metric, such as McCabe, Halstead, or even the C\&K, cannot offer any new information about maintenance if their correlation with size is neutralized. Size has been acknowledged as the predominant code metric even by the studies that support the usefulness of other code metrics~\cite{Spadini:2018, Johnson:2019}. Even many of the code smells, or antipatterns, were found to be directly associated with size~\cite{Khomh:2012}. 
It is no exaggeration to state that size is \emph{``The One Metric to rule them all''}.

One of the primary goals of code metric studies is to provide guidelines about metrics thresholds for maintainability~\cite{DBLP:conf/scam/AnicheTZDG16, Fontana:2015, Alves:2010}. 
And yet, despite being the most important metric, there is no systematic study that provides guidelines about the relationship between size and future maintainability.
In this paper we focus on size thresholds at the method-level granularity. Method level is the most desirable to the developers, because class/file, or module-level granularities are often too coarse to be practically useful~\cite{grund:2021, Shihab:2012,Giger:2012,PASCARELLA:2020}. 

\section{Methodology}
This section describes the methodology of this paper.

\begin{table*}[t]
\begin{center}
    \caption{Description of the top and the bottom five projects (ranked by \# methods) and their method-level statistics. To save space, the full table is shared with our dataset (\texttt{stats/table-1.pdf}). In total, 785,606 Java methods were extracted. 
    Getters and setters and methods that are younger than 2 years are excluded.
    Snapshot SHAs are included to aid reproducibility. 
    }
    \label{tbl:subjectsTable}
    
    \begin{tabular}{ l r r r r r r}
        \toprule
        \textbf{Repository} & \textbf{\# Methods} &\textbf{\# Revisions (avg)}& \textbf{\# Getters-Setters} & \textbf{ Age $\ge$ 2 Years} & \textbf{\# Used Methods}& \textbf{Snapshot}\\
        \midrule
hadoop & 70,524 & 1.8 & 11,219 & 63,378 & 53,729 & \texttt{4c5cd7} \\
elasticsearch & 63,253 & 3.7 & 8,875 & 39,085 & 34,641 & \texttt{92be38} \\
flink & 38,482 & 1.9 & 6,790 & 19,967 & 15,792 & \texttt{261e72} \\
presto & 37,639 & 1.9 & 8,520 & 26,785 & 20,766 & \texttt{bb20eb} \\
lucene-solr & 37,600 & 1.6 & 5,625 & 29,287 & 24,641 & \texttt{b457c2} \\

... & ... & ... & ... & ... & ... & \texttt{...} \\
mockito & 1,526 & 4.2 & 144 & 1,221 & 1,113 & \texttt{077562} \\
cucumber-jvm & 1,222 & 2.7 & 170 & 435 & 374 & \texttt{b57b92} \\
commons-io & 1,149 & 3.1 & 85 & 952 & 884 & \texttt{11f0ab} \\
vraptor4 & 934 & 1.7 & 132 & 933 & 801 & \texttt{593ce9} \\
junit4 & 879 & 3.1 & 56 & 856 & 801 & \texttt{50a285} \\

\hline
\textbf{Total} & \textbf{785,606} & &  & & \textbf{520,874} &   \\

    \bottomrule
    \end{tabular}
    \end{center}
    \vspace{-2mm}
\end{table*}
\subsection{Project Selection}
Code metrics research relies on aggregated~\cite{Gil:2017, Spadini:2018, PASCARELLA:2020} or individual project analysis~\cite{Shin:2011, Zhou:2010, Kafura:1987, Romano:2011}. In aggregated approaches, measurements from a set of selected projects are merged together to produce an observation (e.g., a correlation coefficient between McCabe and change proneness). We argue that this approach is inaccurate. For example, the change-proneness of a method could be influenced by factors that are project dependent, such as the number of active contributors, their commit patterns, code review policy, and the number of releases influences how a software evolves~\cite{Matter:2009, Herzig:2013, WANG:2019}. 
Additionally, aggregated analysis results can be heavily influenced by a few large projects, making observations from smaller projects unnoticeable. 
A more accurate approach is to analyze each project separately to see if an observation is common across projects. 
This approach, however, can be criticized for selection bias~\cite{RADJENOVIC:2013, Gil:2017}. 

We addressed the selection bias threat by selecting all the GitHub projects used by five different previous studies (specifically from~\cite{Ray:2016,Gil:2017,Palomba:2017,Spadini:2018,grund:2021}), resulting in 49 open-source Java projects. 
Table~\ref{tbl:subjectsTable} summarizes the dataset. 
The dataset contains projects from different domains with varied number of methods and varied average number of revisions per method. 
This broad selection of a large number of  projects increases the generalizability of the observations about code metrics and maintenance we can make. 
One thing to note is the low average number of method revisions in Table~\ref{tbl:subjectsTable}. This is because only a small subset of methods undergo many revisions, which increases the utility of approaches that can detect these methods:
if we can detect and improve those few high-churn methods early, we can meaningfully reduce future maintenance effort.

\subsection{Data Collection}
To understand the relationship between method size and change-/bug-proneness, we need method-level evolution history.
We used the state-of-the-art history tracing tool CodeShovel~\cite{grund:2021,Grund:tool}, which returns the complete change history of a given method.
CodeShovel also collects information about when a method was modified, who modified it, what was modified, and why the modification took place (i.e., commit message). 
CodeShovel is robust to method or file rename operations while uncovering a method's history. 
It produces accurate change histories for 90\% methods (with 97\% accuracy for method changes).   

\subsection{Maintenance Effort Indicators} 
User studies are a widely adopted approach~\cite{Abid:2019,Hofmeister:2017,Scalabrino:2016,Buse:2010,Scalabrino:2017,Antinyan:2017,Bauer:2019,Kafura:1987,Curtis:1979,Darcy:2005} for understanding maintenance effort.
The maintenance effort of a code fragment can be understood by the cognitive effort and time required to understand the code by a selected group of users. 
The outcome of user studies, however, are often not generalizable~\cite{Brittain:1982}, and contradictory conclusions can be made by comparing two independent studies~\cite{Johnson:2019, Scalabrino:2017}. 
We have mitigated this threat by using change- and bug-proneness as the maintenance indicators as have been widely used in previous maintenance studies (e.g.~\cite{Giger:2012, Rahman:2017, CATOLINO:2018, Gil:2017, Romano:2011, Shihab:2012}).

\subsubsection{Change-proneness}
A maintainable code fragment (a Java method in our case) should be less change-prone: it should not require too many revisions, and none of its changes should be too big. Supporting this hypothesis, we opted for the following four sub-indicators to understand how change-prone a Java method is.

\textbf{\#Revisions:}
The number of revisions, a common maintenance indicator~\cite{Antinyan:2014,Monden:2002,Shin:2011,Antinyan:2015}, is the total number of times a method has been modified,
regardless of the type of modification---corrective, adaptive, preventive, or perfective~\cite{Darcy:2005}. 

\textbf{ChangeSize:} The number of revisions can be misleading, because the effort required to change two Java methods with the same number of revisions might not be the same. Also, number of revisions can be impacted by developers' commit habit~\cite{WANG:2019}.
One alternative is to use the sum of the change sizes (i.e., \texttt{git diff})~\cite{Scholtes:2016,Shin:2011} considering all the revisions the method underwent. 

\textbf{AdditionOnly:} Does adding new lines require the same effort to delete existing lines? Probably not. Therefore, following prior work~\cite{Shin:2011}, we also use the sum of the number of \emph{added lines only} as an indicator of change-proneness. 

\textbf{EditDistance:} The number of added, and/or deleted lines (as considered in \texttt{git diff}) can be impacted by an individual's coding style. It also does not differentiate between modifying a large line and a small line. We therefore used Levenshtein edit distance~\cite{levenshtein:1966} as our final change-proneness indicator, as recommended by some prior studies~\cite{bigbangs:2019,Scalabrino:2017,Scholtes:2016}.
Levenshtein edit distance counts the number of characters that are added, deleted, or modified, to convert one source code version into another.
\subsubsection{Bug-proneness}
We consider a method as \emph{bug-prone}, if it is associated with a bug-fixing commit. In accordance with earlier stdies~\cite{Mocku:2000, Ray:2016}, a commit is considered as bug-fix if the its commit message contains at least one bug related keyword:  \emph{error}, \emph{bug}, \emph{fixes}, \emph{fixing}, \emph{fix}, \emph{fixed}, \emph{mistake}, \emph{incorrect}, \emph{fault}, \emph{defect}, \emph{flaw}. The level of bug-proneness of a method is the number of bug-fix commits associated with the method.
\subsection{Size Calculation}
\label{size:calculation}
Size is commonly referred to as the source lines of code (SLOC). Some previous studies were explicit about how they calculated SLOC (e.g.,~\cite{Landman:2014, Antinyan:2014, PASCARELLA:2020}), but some studies were not (e.g.,~\cite{Shin:2011, Gil:2017}). 
We calculated SLOC in three different ways, to evaluate if how size is calculated has any impact on its relationship with the maintenance indicators, such as change- and bug-proneness.
 
 \textbf{SLOCStandard} source lines of code without comments and blank lines. This form of SLOC is widely used (e.g.,~\cite{Landman:2014,Ralph:2018}). 
    
\textbf{SLOCAsItIs} source lines of code with comments and blank lines.
    
\textbf{SLOCPretty} The same code snippet, when written by different developers, may produce different SLOC values. By using the PrettyPrinter function of the javaparser~\cite{javaparser}, we convert every source code to a common format producing the same SLOC value for the variants of a code snippet.   

\subsection{Preprocessing}
\label{preprocessing}
\textbf{Getters/setters:} Getters and setters, also known as accessor methods, are often generated automatically and create noise in code metric studies~\cite{Heitlager:2007, Fontana:2015, Liu:2018}. Similar to earlier studies~\cite{Fontana:2015, Abid:2019}, we filtered them out from our dataset, before calculating any code metrics. 
If a method name starts with \texttt{get}, has non-void return type, and does not have any parameter, then we identify this method as a \emph{getter}. Similarly, if a method name starts with \texttt{set}, has one parameter and a void return type, then we label this method as a \emph{setter}. Table~\ref{tbl:subjectsTable} shows the number of getters/setters for each project. 

\textbf{Controlling for Age:} The Kendall's $\tau$ correlation coefficient between ages of methods and their number of revisions is 0.2 in our aggregated dataset. When considered individually, some projects show very high correlation (e.g., 0.51 for the checkstyle project). Therefore, when analyzing the relationship between SLOC and maintenance indicators, we should not include  methods with different ages.
Generally speaking, a ten year old method will have more chances to change than a newly introduced method, regardless of their code complexity.  
To perform age normalization, we followed a two step procedure. In step 1, we removed methods that are less than two years old. This does not, however, solve the problem completely, because we may still compare a 2 years old with a 2+$x$ years old method. Therefore, in step 2, we excluded all revisions and bugs that happened after two years of a method's life. If we set the threshold more than two years, we lose many methods from our dataset. As presented in Table~\ref{tbl:subjectsTable}, with two year threshold, we still have 520,874 methods for our final analyses. If we set the threshold less than two years, we lose many change and bug information, because the window to observe changes becomes narrower. 

\textbf{Multiple Versions:} A method can have multiple SLOCs while it evolves. Consider a method associated with 5 SLOCs in its lifetime: 20, 50, 50, 50, 22. The method was revised 4 times after its introduction, so the value of \#revisions is 4. To calculate the correlation between SLOC and \#revisions, what SLOC value should represent this method? If we take the introduction SLOC, we will map SLOC 20 to 4 revisions, which is inaccurate because the method was revised only once for SLOC 20. We solve this problem by applying a versioning technique, where the above method has three different versions. That way version 1 has SLOC 20 with 1 revision, version 2 has SLOC 50 with 3 revisions, and version 3 has SLOC 22 with 0 revision, because the method did not change afterwards. We apply this technique for other maintenance indicators as well.


\subsection{Statistical Tests}
Throughout this paper, we compare if different distributions are statistically different. We applied the Anderson-Darling normality test~\cite{thode2002testing} to some of the randomly selected distributions and found that many of the distributions are not normally distributed. Therefore, non-parametric tests are selected for our statistical testing. For evaluating if two distributions are different, we use the Wilcoxon rank sum test (and Wilcoxon signed rank test for pairwise comparison).
For calculating the correlation coefficient between two variables, we use the non-parametric Kendall's $\tau$, instead of the Pearson's correlation. 
For calculating effect sizes, we apply Cliff's Delta instead of Cohen's d. In accordance with~\cite{Hess2004RobustCI}, we classify an effect size either as Negligible ($0\le |\delta| < 0.147$), Small ($0.147 \le |\delta| < 0.330$), Medium ($0.330 \le |\delta| < 0.474$), or Large ($0.474 \le |\delta| \le 1$).
These non-parametric tests are widely used in software engineering research~\cite{Inozemtseva:2014, Gil:2017, greenscaler:2019}.  
\section{Results}
In this section, we present our findings to our four research questions. We investigate whether method size is an important factor to understand change- and bug-proneness, what the upper-limit of method's size should be, whether developers should decompose large methods into smaller ones, and if controlling for size controls other code quality factors. 

\subsection{RQ1: Is SLOC correlated with change- and bug-proneness at method-level?}

\begin{figure}[htbp]
\centering
\includegraphics[width=0.45\textwidth ,keepaspectratio]{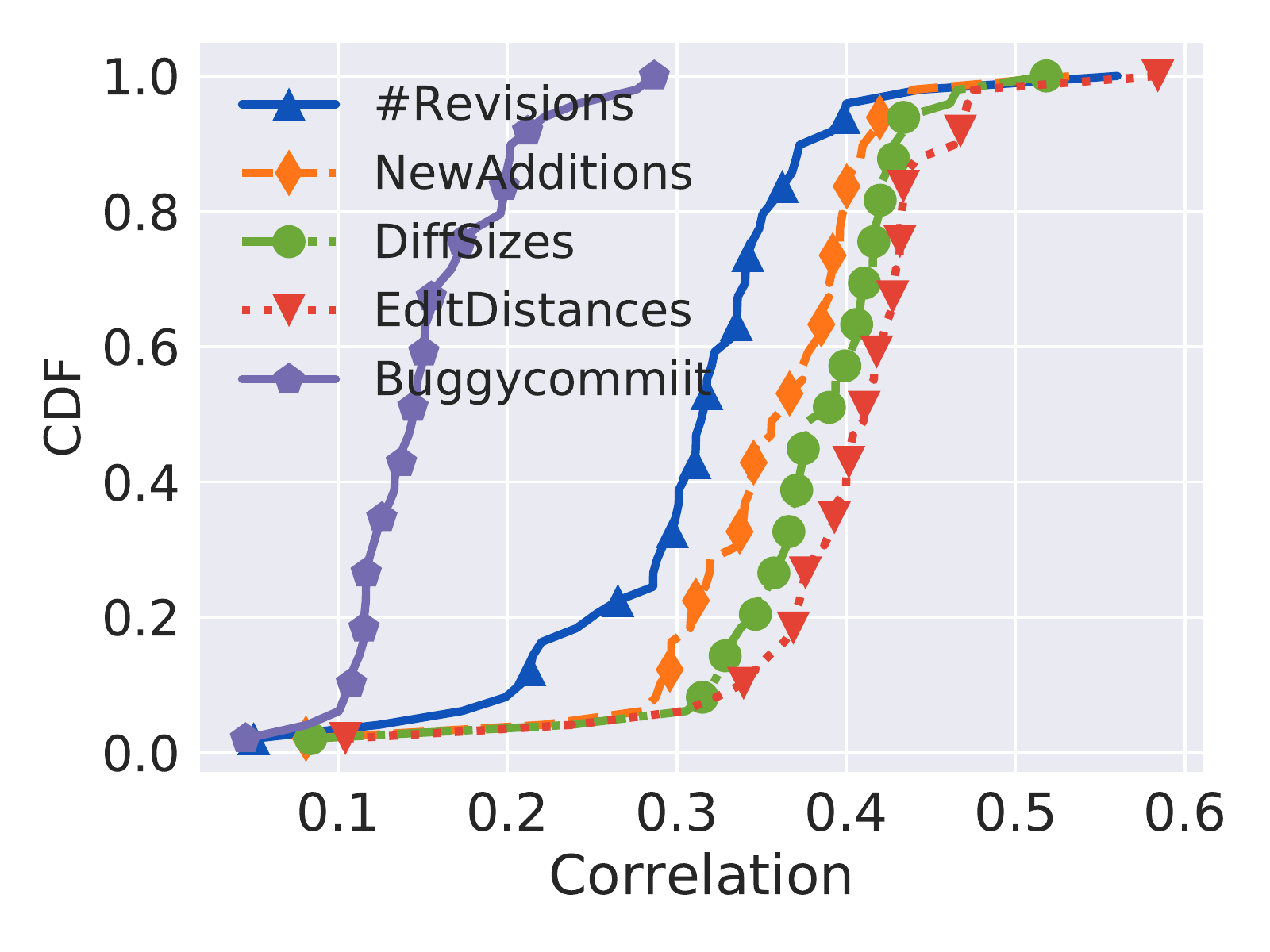}

\caption{Cumulative distribution functions of the correlation between SLOC and the five maintenance indicators. Each line represents 49 points for our 49 different projects. For graph readability, number of markers has been reduced.  
} 
\label{fig:sloc-maintenance}
\end{figure}

Figure~\ref{fig:sloc-maintenance} shows the distribution of Kendall's $\tau$ correlation coefficients between SLOC and the five maintenance indicators. Correlation coefficient for each project is shared with our pubic dataset as a table as well (\texttt{stats/table-2.pdf}). For all projects, the correlation coefficients are positive, indicating method size is an indicator of future maintenance. When compared among the four change-proneness indicators, SLOC performance is generally the lowest for \#revisions, and highest for EditDistance. Correlations between SLOC and bug-proneness (i.e., \#BuggyCommits) are significantly lower than the four change-proneness indicators. This implies that, at the method-level, code metrics are comparatively less helpful for bug prediction than for change prediction. 
This complements the recent findings of Pascarella \textit{et al.}~\cite{PASCARELLA:2020}, who consider method-level bug prediction as an open research problem.   
We argue that it is unrealistic to expect very high correlation between a code metric and maintenance effort. A very high correlation would mean that maintenance effort can be
estimated just by using one code metric. This is unrealistic because there are many factors that influence code maintenance, such as  developer habits~\cite{Terceiro:2010}, application domain and platforms~\cite{Zhang:2013,viggiato:2019}, code clones~\cite{Monden:2002}, software architecture~\cite{DBLP:conf/scam/AnicheTZDG16}, test code quality~\cite{Spadini:2018, DBLP:journals/tse/AthanasiouNVZ14}, and changes in requirements.

\begin{figure}[htbp]
\centering
\includegraphics[width=0.45\textwidth ,keepaspectratio]{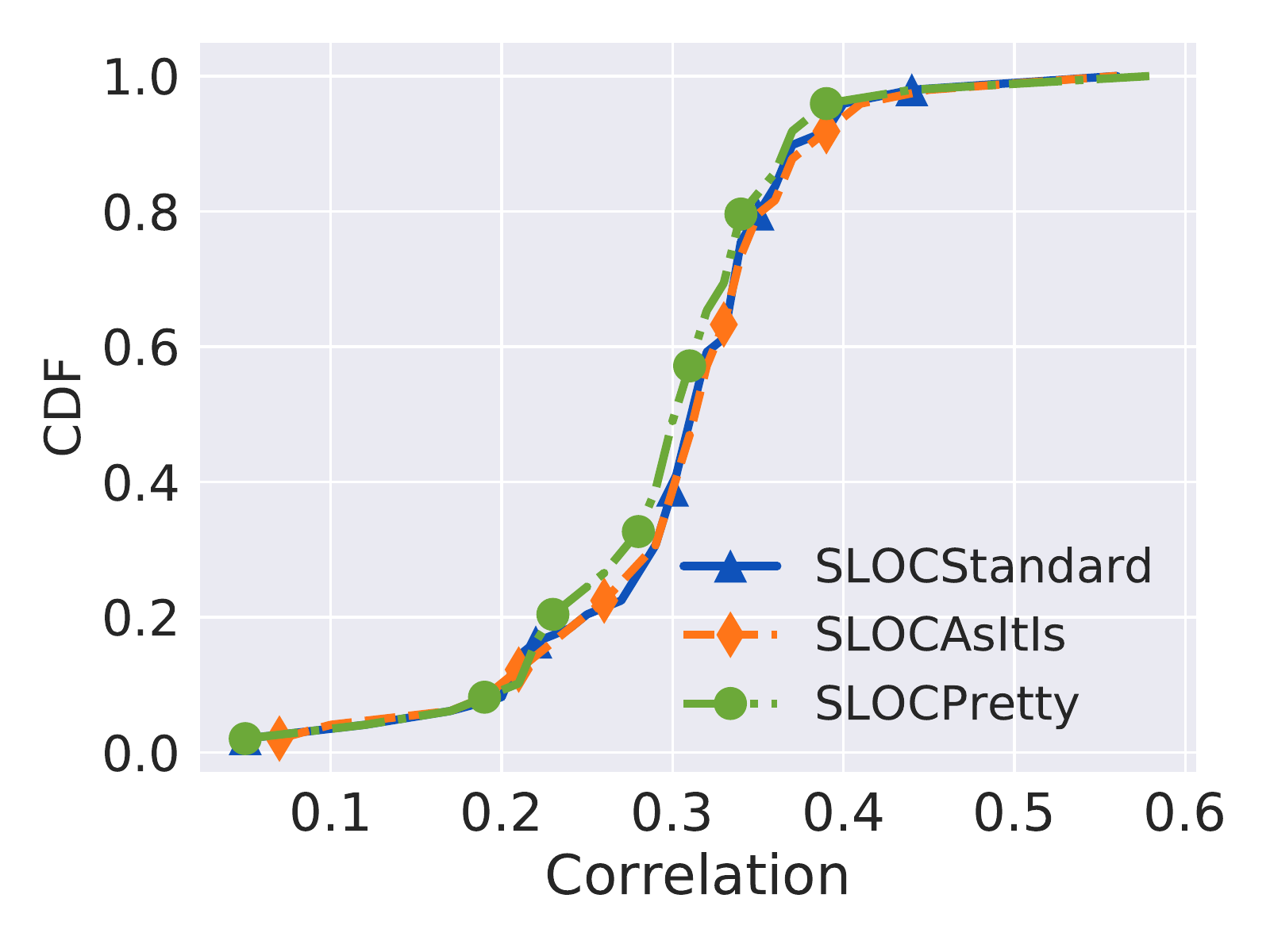}

\caption{Cumulative distribution functions of the correlation between different SLOC types and number of revisions. 
} 
\label{fig:correlation-compare}
\end{figure}

We also observe that SLOC performance is not similar across projects. For the docx4j project, SLOC correlates to change-proneness very weakly (e.g., correlation is only 0.05 with \#revisions) while for the voldemort  project the correlation is 0.56. 
This observation is true for bug-proneness as well. For the cucumber-jvm project, the correlation is only 0.05, but for the xerces2-j, the correlation is 0.29. These observations explain why cross project maintenance prediction is still a challenging problem~\cite{Tantithamthavorn:2018,Zhang:2013}.

\begin{table*}[htbp]
\begin{center}
    \caption{P-values from Wilcoxon signed rank test (i.e., pairwise test). Most of the time SLOCPretty is significantly different than the other two ($p \le 0.05$).  
    }
    \label{tb:type-comparison}

    \begin{tabular}{ c c r r r r r }

        \toprule
        \textbf{Type1} & \textbf{Type2} & \textbf{\# Revisions} & \textbf{\# Additions} & \textbf{DiffSize} & \textbf{EditDistance} & \textbf{Bugs} \\
        \midrule
        SLOCStandard&SLOCAsItIs&0.45&0.78&0.06&0.00&0.02\\
SLOCStandard&SLOCPretty&0.00&0.00&0.00&0.00&0.07\\
SLOCAsItIs&SLOCPretty&0.00&0.00&0.00&0.00&0.01\\

        \bottomrule
    \end{tabular}
    \end{center}
   \vspace{-2mm}
\end{table*}

Unless otherwise mentioned, we have considered SLOC as the source lines of code without comments and blank lines (SLOCStandard), which we found to be the most common in the literature (e.g.,~\cite{Landman:2014,Ralph:2018}). No previous study, however, has investigated if
this is the best size measure to understand maintenance. With \#revisions as the maintenance indicator, Figure~\ref{fig:correlation-compare} shows the performance comparison among the three different SLOC calculations (described in Section~\ref{size:calculation}). Both SLOCStandard and SLOCAsItIs slightly outperform SLOCPretty. According to the Wilcoxon signed rank test (for pairwise testing), the performance between SLOCStandard and SLOCAsItIs are not statistically different ($p > 0.05$). Their performance, however, are statistically different with SLOCPretty ($p \le 0.05$). Table~\ref{tb:type-comparison} shows the results for all the maintenance indicators. 

We also calculated the Cliff's Delta effect sizes among the performance distributions. The differences are always negligible, except for two cases. These two cases are between SLOCPretty and SLOCAsItIs: one for the DiffSize and the other for the EditDistance indicator. The signs of the Cliff's Delta values (+/-), however, suggest that SLOCPretty performs the worst whereas SLOCStandard performs the best, although with negligible effect sizes.

\begin{hassanbox}
\textbf{Summary:} SLOC correlates with change- and bug-proneness at the method-level granularity. SLOC's correlation with bug-proneness is generally lower than change-proneness. The best way to calculate SLOC is to calculate source lines of code without comments and blank lines. 
\end{hassanbox}
\subsection{RQ2: What is the maintainable method size?}
Our previous analysis shows that method SLOC is positively correlated with maintenance effort. This means developers should try to keep their method size as small as possible. \emph{But how small should methods be?} Suggesting a very low upper-bound, such as \emph{`keep your method size within 5 SLOC'}, can be impractical. In order to find a method size limit that is realistic and maintainable at the same time, we follow two steps. In step 1, we find a realistic method size limit by studying all the methods from the 49 Java projects in our dataset. In step 2, we empirically evaluate if this limit can indeed reduce maintenance effort by reducing change- and bug-proneness. 
\subsubsection{A realistic upper bound of method size}
Alves \textit{et al.}~\cite{Alves:2010} proposed a 6-steps systematic approach to find thresholds for any given code metric, such as McCabe. The methodology does not depend on intuition or expert opinion and is   
resilient to outlier projects. From the source code of a set of software projects, the methodology can automatically derive thresholds of a code metric in four sizes: small, medium, large, and very large. We apply this methodology to find these four thresholds for SLOC at the method-level granularity for our dataset of 49 projects. Although we refer to~\cite{Alves:2010, Fontana:2015, Yamashita:2016} for more details, for reproducibility, we here describe the exact process followed. To have better understanding, let us consider two projects each containing three methods.  For the first project ($P^1$), SLOCs of the methods are 10, 20, and 10, whereas for the second project ($P^2$), the SLOCs are 20, 10, and 20.

\textbf{Step 1} For each method $M^{m}$ in project $P^{1}$, we calculate its SLOC: $P^1_{M^{m}_{sloc}}$. 
So for the first project, we have: \\\\ $P^1_{M^{1}_{sloc}}= 10, P^1_{M^{2}_{sloc}}= 20, P^1_{M^{3}_{sloc}}= 10$\\

\textbf{Step 2} For each $M^{m}$ in $P^1$, we calculate its weight ratio, $P^1_{M^{m}_{weight}} = P^1_{M^{m}_{sloc}} / \sum_{k=1}^{n}{P^1_{M^{k}_{sloc}}}$, where n is the number of methods in project $P^{1}$. Therefore: \\\\ $P^1_{M^{1}_{weight}} = P^1_{M^{1}_{sloc}}/(P^1_{M^{1}_{sloc}}+P^1_{M^{2}_{sloc}}+P^1_{M^{3}_{sloc}})=0.25$ \\
\\\\Similarly, $P^1_{M^{2}_{weight}}=0.5, P^1_{M^{3}_{weight}}=0.25$ \\  

\textbf{Step 3}  All methods with identical SLOC are grouped together. For example, methods $P^p_{M^i},P^p_{M^j}$ and $P^p_{M^k}$ are grouped together if\\\\ $P^p_{M^{i}_{sloc}}=P^p_{M^{j}_{sloc}}=P^p_{M^{k}_{sloc}}$.\\\\
The entity aggregation of this method group \\\\ $ = P^p_{M^{i}_{weight}} + P^p_{M^{j}_{weight}}+ P^p_{M^{k}_{weight}}$.\\\\
In our case with $P^1$, the entity aggregation for the first method group with SLOC 10 \\\\$ = P^1_{M^{1}_{weight}} + P^1_{M^{3}_{weight}} = 0.5$, \\\\ and for the second method group with SLOC 20$ = P^1_{M^{2}_{weight}} = 0.5$ \\\\
Each group's SLOC value is added to a list called $slocs$. The aggregated value is added to another list called $aggr$. So, we have $slocs = [10, 20]$, and $aggr = [0.5, 0.5]$

\textbf{Step 4} For each  entry $i$ in $aggr$, we calculate the normalized entity aggregation $aggr^i / \phi$, where $\phi$ is the number of projects. We add this normalized value to another list called $norms$. Therefore, $norms = [0.25, 0.25]$ 

\textbf{Step 5} For each project $p = 1,2,.., \phi$, we repeat step 1 to 4. This produces $\phi$ different lists of $slocs$ (called $list\_slocs$) and $norms$ (called $list\_norms$). In our case, $list\_slocs = [[10,20],[20, 10]]$, and $list\_norms = [[0.25,0.25], [0.4,0.1]]$.
We used this two lists, to calculate another two lists called $x\_axis$ and $y\_axis$, using an algorithm suggested by Alves \textit{et al.}~\cite{Alves:2010}. For reproducibility, the algorithm is shared with our dataset package (\texttt{algorithm/algorithm.png}).





\begin{figure}[htbp]
\centering
\includegraphics[width=0.45\textwidth ,keepaspectratio]{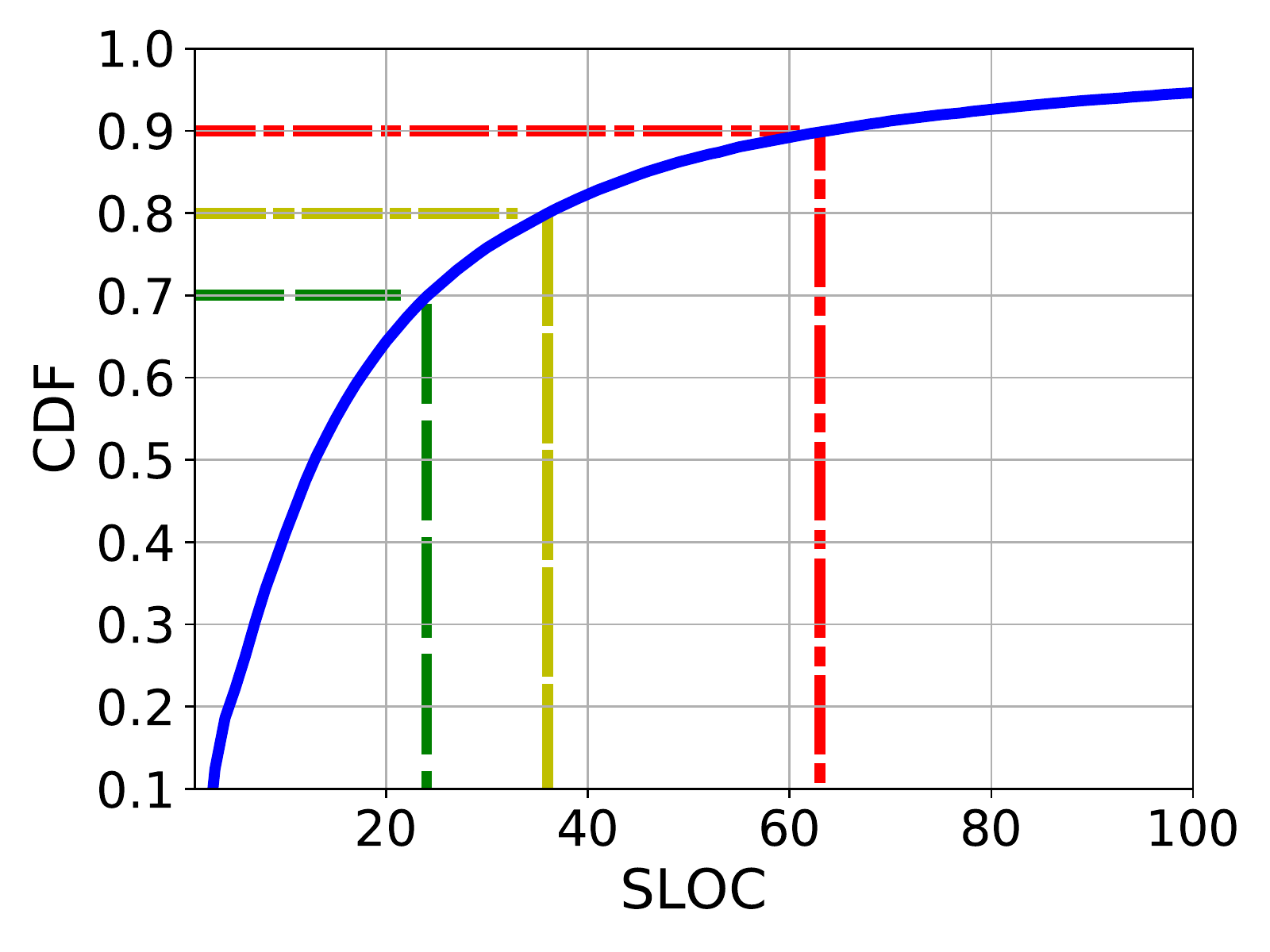}

\caption{Cumulative line chart from Alves \textit{et al.}'s approach.
} 
\label{fig:thresholds-cutoff}
\end{figure}

\textbf{Step 6} From the algorithm, $x\_axis = [10, 20]$, $y\_axis = [0.35, 0.65]$. We now sort $x\_axis$ in ascending order (and $y\_axis$ accordingly) to produce a cumulative line chart. Figure ~\ref{fig:thresholds-cutoff} shows the cumulative line chart after following all the steps for all of the 49 projects. From the line chart, we get three critical values to categorize a method either as small, or medium, or large, or very large. According to the Alves \textit{et al.}'s approach, the first critical value in Figure~\ref{fig:thresholds-cutoff} is 24 (from the x-axis), because it covers 70\% of the y-axis. 
The second and the third values are  36 and 63, because they cover 80\% and 90\% of the y-axis respectively. 

Based on these three critical values, Table~\ref{tb:size} explains how to determine if a method is small, medium, large, or very large.  For example, if a method is 24 SLOC or fewer, it is a small method while a method with 64 SLOC or more is a very large method. We conclude that 24 SLOC is not only small but also realistic, because majority of the methods can be written within 24 SLOC (Figure~\ref{fig:thresholds-cutoff}). This observation is true even without getters/setters, because we have removed them from our dataset, as explained in Section~\ref{preprocessing}.

\begin{table}[htbp]
\begin{center}
    \caption{Determining the size of a method after applying Alves \textit{et al.}'s approach in our dataset of 49 projects. 
    }
    \label{tb:size}

    \begin{tabular}{ c r r}

        \toprule
        \textbf{Size} & \textbf{Lower-bound} & \textbf{Upper-bound} \\
        \midrule
        Small & -- & 24 \\
        Medium & 25 & 36 \\
        Large & 37 & 63 \\
        Very Large & 64 & -- \\
        \bottomrule
    \end{tabular}
    \end{center}
    \vspace{-2mm}
\end{table}


\begin{figure*}[htbp]
\centering
\includegraphics[width=0.9\textwidth ,keepaspectratio]{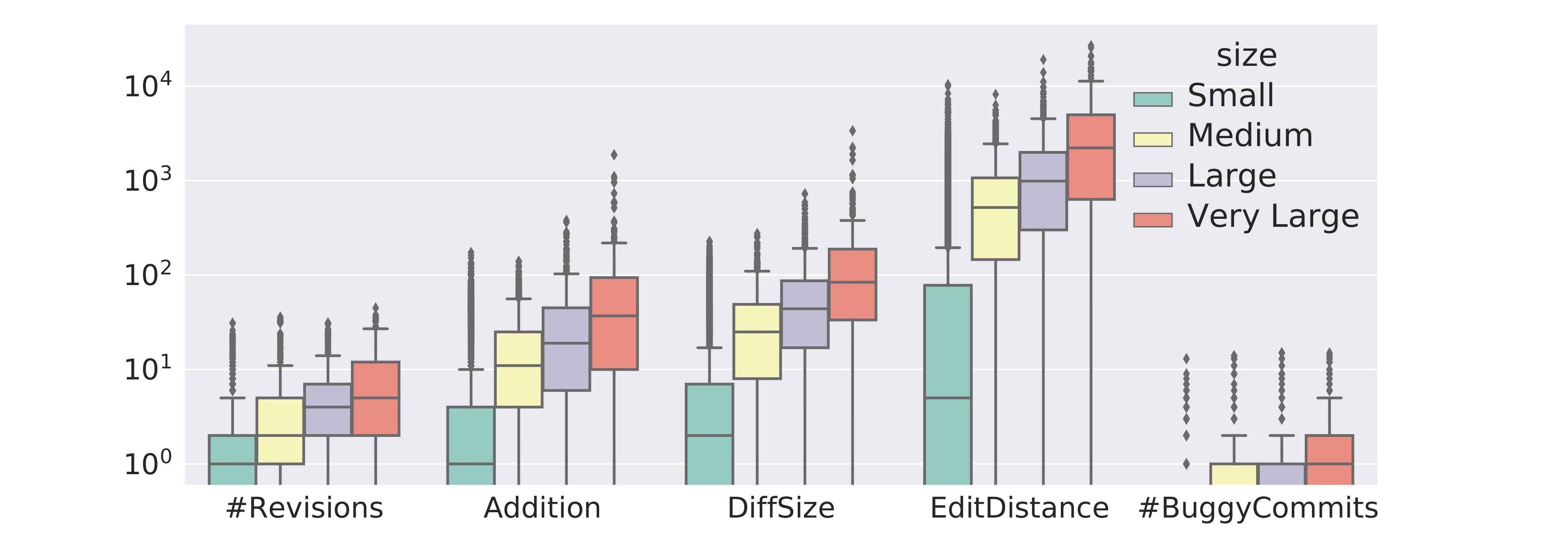}

\caption{For the elasticsearch project, this graph compares the distributions of different maintenance indicators after grouping them according to the four different method sizes.    
} 
\label{fig:sample}
\end{figure*}

\begin{table*}[htbp]

\begin{center}
    \caption{Wilcoxon rank sum test to show the percentage of cases where the SLOC categorization exhibit statistically significantly different maintenance effort.    
    }
 \label{tbl:percent-projects}

    \begin{tabular}{ l r r r r r }
        \toprule
        \textbf{Sample Projects} & \textbf{\# Revisions} & \textbf{Addition} & \textbf{DiffSize} & \textbf{EditDistance} & \textbf{\#BuggyCommits} \\
        \midrule
        \textbf{All} & 82.98\% & 86.52\%& 87.23\% & 87.94\% & 77.30\%\\
        \textbf{Top 20} & 96.67\% & 96.67\% & 96.67\% & 96.67\% & 91.67\%\\
        \bottomrule
    \end{tabular}
    \end{center}
\end{table*}

\begin{table*}[htbp]
\begin{center}
    \caption{Cliff's Delta Effect sizes after comparing methods in different size categories (e.g., Small-Medium indicates comparison between small and medium size methods). \texttt{N} refers to Negligble, \texttt{S} refers to Small, \texttt{M} refers to Medium, and \texttt{L} refers to Large effect size. For example, the first column of the first row compares the \#revisions distributions between small and medium size methods and shows the percent of differences with negligible, small, medium, and large effect sizes.}
     \label{tbl:effect-sizes}
    \begin{tabular}{ l| c| c| c| c| c| c| c| c|c|c|c|c|c }
        \toprule
        & \multicolumn{4}{c|}{\textbf{Small-Medium}} & \multicolumn{4}{c|}{\textbf{Medium-Large}} & \multicolumn{4}{c|}{\textbf{Large-Very Large}}\\
        
        \textbf{Indicator} & \texttt{N}&\texttt{S}&\texttt{M}&\texttt{L} & \texttt{N}&\texttt{S}&\texttt{M}&\texttt{L} & \texttt{N}&\texttt{S}&\texttt{M}&\texttt{L} \\
        \midrule
        
\textbf{\# Revisions} &0.00&10.20&55.10&34.69&35.42&58.33&2.08&4.17&29.55&59.09&6.82&4.55 \\ 
\textbf{Addition} &0.00&6.12&20.41&73.47&29.17&60.42&8.33&2.08&27.27&59.09&11.36&2.27 \\
\textbf{DiffSize}&0.00&6.12&6.12&87.76&22.92&66.67&4.17&6.25&20.45&52.27&22.73&4.55 \\
\textbf{EditDistance}&0.00&4.08&6.12&89.8&18.75&60.42&16.67&4.17&22.73&45.45&25.0&6.82 \\
\textbf{\#BuggyCommits} &38.78&57.14&4.08&0.00&70.83&22.92&6.25&0.00&52.27&43.18&2.27&2.27\\        \bottomrule
    \end{tabular}
    \end{center}
    \vspace{-2mm}
\end{table*}

\subsubsection{Evaluating the upper bound for maintenance}
We found that 24 SLOC is a reasonable upper bound of method size, but we need to evaluate if this bound makes any real difference with the other method sizes (medium, large, and very large) when compared against change- and bug-proneness. To understand how we evaluate this for the 49 different projects, let us consider Figure~\ref{fig:sample} as a demonstration for the elasticsearch project. In the first group of four box-plots, we compare the \#revisions distributions, where each box-plot contains all the \#revisions for all the methods with a specific size (e.g., the first box-plot shows the \#revisions distribution for all the small methods with SLOC $\le 24$). Clearly, \#revisions increases with the increase in method sizes. 

To observe if these differences in \#revisions are statistically different and what are the effect sizes of those differences, we apply three comparisons. These comparisons are between distributions of \#revisions for i) small methods and medium methods, ii) medium methods and large methods, and iii) large methods and very large methods. According to the Wilcoxon rank sum test, in all the three cases, the differences are statistically different ($p \le 0.05$). The Cliff's Delta effect size is \emph{large} between small and medium methods, and \emph{small} for the other two cases. These observations are true for the other four maintenance indicators as well. 

\emph{Are these observations generalizable?} According to our comparison approach, for each indicator (e.g., \#revisions) we have a total of 147 comparisons (3 comparisons per project $\times$ 49 projects). Table~\ref{tbl:percent-projects} shows that for \#revisions, in 82.98\% of cases the comparisons are statistically different. This percent is the highest for the EditDistance, and the lowest for the \#BuggyCommits. Many of our projects, however, do not have enough methods in each size category to make a reliable comparison. Therefore, we also consider only the top 20 projects, after ranking all the projects based on their number of methods. 
Table~\ref{tbl:percent-projects} shows that the percent of cases with significant statistical differences are much higher for the top 20 projects. 

We also calculated the effect sizes of the differences, as presented and described in Table~\ref{tbl:effect-sizes}. For all the four change-proneness indicators, no project exhibits negligible effect size when compared small methods with medium size methods. In fact, most of the cases are within medium to large effect sizes. For example, for \#revisions, 55.1\% of the effect sizes are medium, and 34.69\% are large. This observation, however, is not true when compared for the other size categories: \textbf{Medium-Large}, and \textbf{Large-Very Large}. For these two groups, most of the effect sizes are within negligible and small. These observations suggest that converting a medium method to a large method, or a large method to a very large method are not as harmful as converting a small method to medium method. 
Consistent with the observation from RQ1, for the bug-proneness indicator (i.e., \#BuggyCommits), the observation is different than the observations with the four change-proneness indicators. Most of the differences are now within negligible and small effect sizes. And yet, the difference in the \textbf{Small-Medium} group is more meaningful than the others, because it has the lowest percent (38.78\%) of negligible differences. 
We conclude that, a threshold $<$ 24 would be less achievable (Figure~\ref{fig:thresholds-cutoff}), whereas a threshold $>$ 24 would incur significantly higher maintenance effort (Table~\ref{tbl:effect-sizes}).

\begin{hassanbox}
\textbf{Summary:} Methods of 24 SLOC or fewer exhibit less change- and bug-proneness. Therefore, developers should strive to keep their methods within 24 SLOC.  
\end{hassanbox}

\subsection{RQ3: Should developers decompose large methods?} 
Some methods are inherently large, due to the complicated tasks they implement. 
Would refactoring these complex methods into smaller methods ($\le 24 $ SLOC) actually improve maintainability, or just distribute it?
This is important, because decomposing large methods would increase dependency and coupling score (more fan-in and fan-out) of a software project, which has been shown harmful for software maintenance~\cite{Mo:2016}. 
To investigate this issue, we would require two versions of each software project. In one version, method size was not controlled, and in the other each method with SLOC $>24$ was decomposed. Unfortunately, such a project does not exist. Therefore, \emph{we consider that any group of small methods that can easily be merged to a larger method are the result of a large method decomposition.} 
We then compare the change- and bug-proneness of individual methods having SLOC $ > 24$ with the summation of change- and bug-proneness of small methods (SLOC $\le 24$) that are candidate for merging. 

\emph{Rules for merging: any method \texttt{B} is a candidate to merge with any method \texttt{A}, if and only if i) \texttt{B} is called only by method \texttt{A}, ii) both \texttt{A} and \texttt{B} have SLOC $\le 24$, and iii) the SLOC limit for \texttt{A} will not apply if \texttt{A} is a result of previous merging.} 

\begin{figure}[htbp]
\centering
\includegraphics[width=0.45\textwidth ,keepaspectratio]{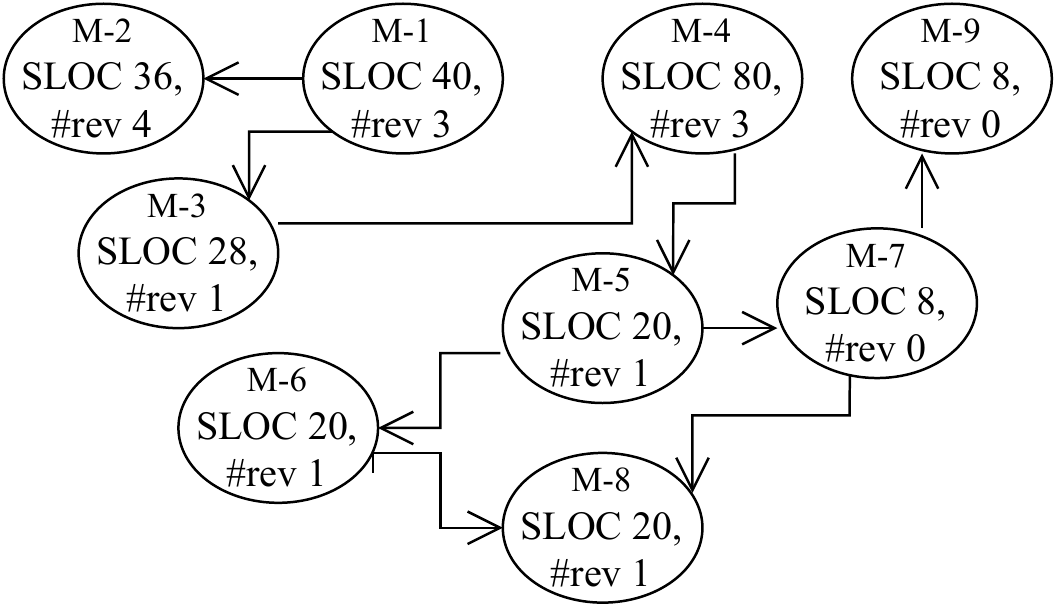}

\caption{Example of a call graph.} 
\label{fig:call-graph}
\end{figure}

While the first two rules are intuitive, \emph{rule iii} is to produce arbitrarily large SLOC so that we can compare with similar sized individual methods. To better understand the procedure, let us consider the call graph in Figure~\ref{fig:call-graph}. For simplicity, we show \#revisions as a method's attribute, although the procedure is the same for the other maintenance indicators. Here, methods M1, M2, M3, and M4 are candidate for decomposition, because their SLOCs are larger than 24. On the other hand, with our \emph{rules for merging}, method M6 is a candidate to merge with M5, producing a total SLOC 40, and total \#revisions 2. Similarly, we can merge M7 with M5, and M9 with M7. We can not merge M5 with M4, because M4 violates \emph{rule ii} (SLOC $>24$). M8 can not be merged with any methods because it is called by more than one methods (\emph{rule i}). We, however, can merge M9 with M7 and then the resulted method with M5, because none of this methods has SLOC $>24$ when considered individually (\emph{rule iii}). Table~\ref{tb:merge} shows the two groups of methods with their SLOCs and \#revisions. The \textit{Selected} column is described later. 


\begin{table}[htbp]
\begin{center}
    \caption{Grouping methods from Figure \ref{fig:call-graph} either as individual or merged methods according to our \emph{rules for merging}. 
    }
    \label{tb:merge}

    \begin{tabular}{ c r r r}

        \toprule
        \textbf{Individual Method} & \textbf{SLOC} & \textbf{\#revisions} & \textbf{Selected}\\
        \midrule
         M1 & 40 & 3 & Yes\\
         M2 & 36 & 4 & Yes\\
         M3 & 28 & 1 & Yes\\
         M4 & 80 & 3 & No\\\hline
         \textbf{Merged Methods} & \textbf{Total SLOC} & \textbf{Total \#revisions} & \textbf{Selected} \\\hline
         M5, M6 & 40 & 2 & Yes\\
         M5, M7 & 28 & 1 & Yes\\
         M7, M9 & 16 & 0 & No\\
         M5, M7, M9 & 36 & 1 & Yes\\
        \bottomrule
    \end{tabular}
    \end{center}
\end{table}

\begin{figure}[htbp]
\centering
\includegraphics[width=0.45\textwidth ,keepaspectratio]{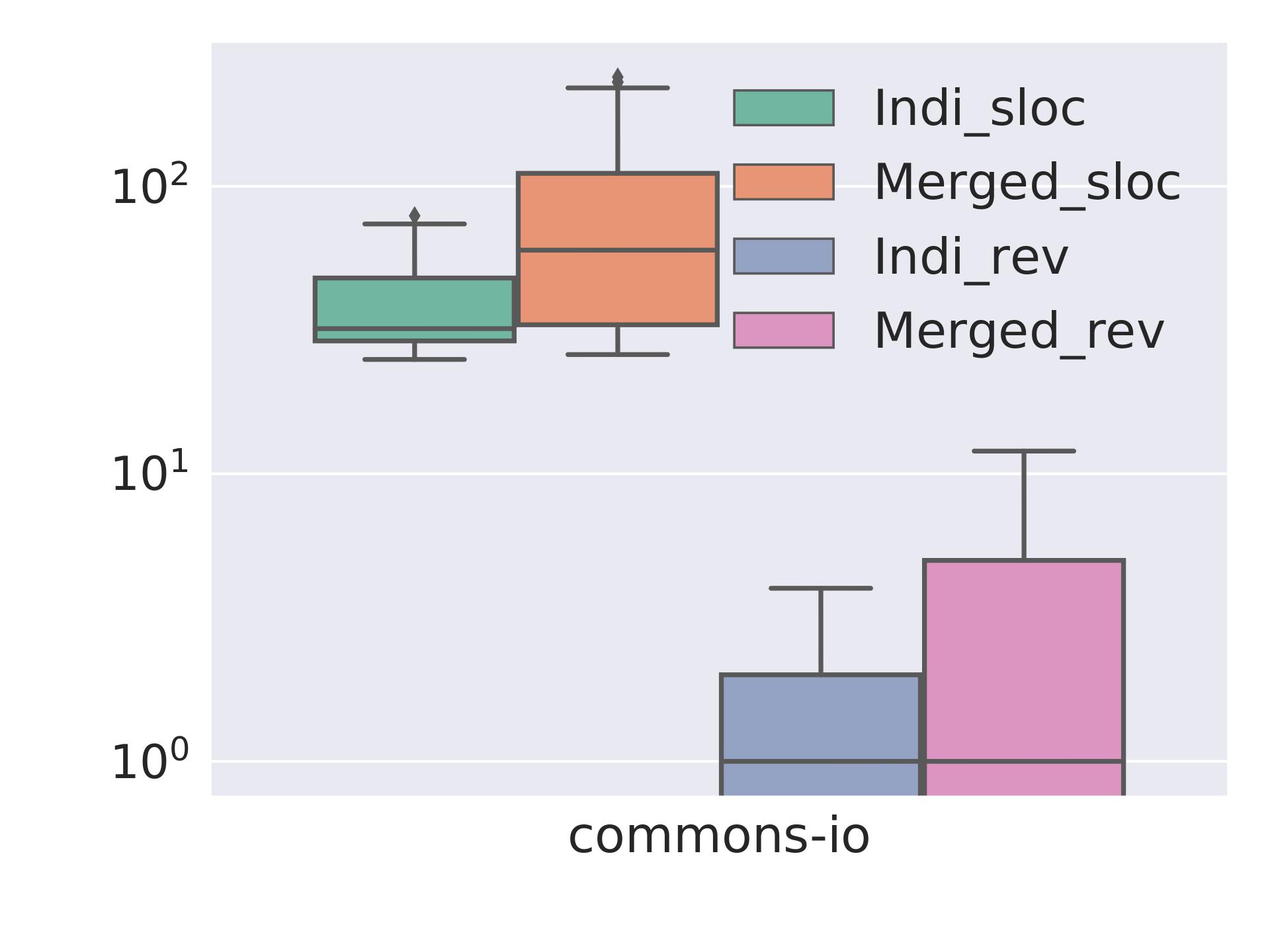}

\caption{ Comparing \#revisions distributions between the two groups for the commons-io project. 
Ind$\sim$Individual, rev$\sim$\#revisions. Here, SLOC distributions are not similar, leading to inaccurate comparison of maintenance indicators.   
} 
\label{fig:problem}
\end{figure}

From Table~\ref{tb:merge}, we can now compare the two distributions of \#revisions for deciding which group, between the individual and merged methods, undergoes less \#revisions. Unfortunately, the SLOC distributions of this two groups are different. Therefore, the comparison will not be accurate, as SLOC influences all the maintenance indicators (RQ1, and RQ2). This problem is depicted in Figure~\ref{fig:problem} for the commons-io project. The second group of box-plots suggests that \#revisions is higher for merged methods (i.e., decomposition is harmful). The first group of box-plots, however, shows that merged methods SLOC distribution is much higher than individual methods, leading to inaccurate comparison between the two distributions of \#revisions. We solve this problem by removing entries that do not have a common SLOC, as presented in the \emph{selected} column in Table~\ref{tb:merge}. For example, the M4 method is not selected because its SLOC (80) does not match with any other methods in the second group of merged methods. For the same reason, the merged method M7, M9 is not selected. After removing these unmatched entries, we can now accurately compare the maintenance indicators of the individual and the merged methods. 

\begin{table*}[htbp]
\begin{center}
    \caption{Distribution of different maintenance indicators of individual methods and merged methods are compared using the Cliff's Delta. Effect sizes are N$\sim$Negligible, and S$\sim$Small. A deep-blue cell means merged method has lower value for that specific indicator---suggesting decomposition is less expensive. For example, for hadoop, the cell for \#revisions is deep-blue and the value is \textit{N}. It means, according to the Cliff's Delta, \#revisions is less for the merged methods with negligible effect size.      
    }
    \label{tb:colored-table}
            \begin{tabular}{ l r r r r r r }
        \toprule
        \textbf{Project} & \textbf{\# Samples} & \textbf{\# Revisions} & \textbf{\# Additions} & \textbf{DiffSize} & \textbf{EditDistance} & \textbf{Bugs} \\
        \midrule
        \textbf{hadoop} & 1304 &   \cellcolor{teal}\textit{N} & \cellcolor{teal}\textit{N}& \cellcolor{teal}\textit{N} &\cellcolor{teal}\textit{N}  & \cellcolor{purple}\textit{N}\\
        
             \textbf{weka} & 1274 &   \cellcolor{teal}\textit{N} & \cellcolor{teal}\textit{N}& \cellcolor{teal}\textit{N} &\cellcolor{teal}\textit{N}  & \cellcolor{teal}\textit{N}\\

        \textbf{hibernate-orm} & 841 &   \cellcolor{purple}\textit{N} & \cellcolor{teal}\textit{N}& \cellcolor{teal}\textit{N} &\cellcolor{purple}\textit{N}  & \cellcolor{purple}\textit{N}\\

        \textbf{hbase} & 742 &   \cellcolor{teal}\textit{N} & \cellcolor{teal}\textit{N}& \cellcolor{teal}\textit{N} &\cellcolor{teal}\textit{N}  & \cellcolor{teal}\textit{N}\\
    
       \textbf{docx4j} & 498 &   \cellcolor{purple}\textit{N} & \cellcolor{purple}\textit{N}& \cellcolor{purple}\textit{N} &\cellcolor{purple}\textit{N}  & \cellcolor{purple}\textit{N}\\
                        
         \textbf{hazelcast} & 493 &   \cellcolor{teal}\textit{N} & \cellcolor{teal}\textit{S}& \cellcolor{teal}\textit{S} &\cellcolor{teal}\textit{S}  & \cellcolor{purple}\textit{N}\\
         
          \textbf{netty} & 445 &   \cellcolor{purple}\textit{N} & \cellcolor{purple}\textit{N}& \cellcolor{purple}\textit{N} &\cellcolor{purple}\textit{N}  & \cellcolor{purple}\textit{N}\\
          
             \textbf{jgit} & 413 &   \cellcolor{teal}\textit{S} & \cellcolor{teal}\textit{S}& \cellcolor{teal}\textit{S} &\cellcolor{teal}\textit{S}  & \cellcolor{teal}\textit{S}\\
             
        \textbf{wicket} & 412 &   \cellcolor{teal}\textit{N} & \cellcolor{teal}\textit{N}& \cellcolor{teal}\textit{N} &\cellcolor{teal}\textit{N}  & \cellcolor{teal}\textit{N}\\

     \textbf{voldemort} & 184 &   \cellcolor{purple}\textit{N} & \cellcolor{purple}\textit{N}& \cellcolor{purple}\textit{N} &\cellcolor{purple}\textit{N}  & \cellcolor{teal}\textit{N}\\
     
     \textbf{openmrs-core} & 170 &   \cellcolor{teal}\textit{N} & \cellcolor{teal}\textit{N}& \cellcolor{teal}\textit{N} &\cellcolor{teal}\textit{N}  & \cellcolor{teal}\textit{N}\\
            
     \textbf{wildfly} & 163 &   \cellcolor{teal}\textit{N} & \cellcolor{teal}\textit{N}& \cellcolor{teal}\textit{N} &\cellcolor{teal}\textit{N}  & \cellcolor{teal}\textit{N}\\
                    
       \textbf{mongo-java-driver} & 144 &   \cellcolor{teal}\textit{S} & \cellcolor{teal}\textit{S}& \cellcolor{teal}\textit{S} &\cellcolor{teal}\textit{S}  & \cellcolor{teal}\textit{N}\\
       
           \textbf{checkstyle} & 137 &   \cellcolor{purple}\textit{S} & \cellcolor{purple}\textit{S}& \cellcolor{purple}\textit{S} &\cellcolor{purple}\textit{S}  & \cellcolor{teal}\textit{N}\\
              
                 \textbf{javaparser} & 105 &   \cellcolor{teal}\textit{N} & \cellcolor{teal}\textit{N}& \cellcolor{teal}\textit{N} &\cellcolor{teal}\textit{N}  & \cellcolor{teal}\textit{N}\\
                  
                 \textbf{hibernate-search} & 78 &   \cellcolor{teal}\textit{N} & \cellcolor{teal}\textit{N}& \cellcolor{teal}\textit{N} &\cellcolor{teal}\textit{N}  & \cellcolor{teal}\textit{N}\\
                     
          \textbf{spring-boot} & 72 &   \cellcolor{purple}\textit{N} & \cellcolor{purple}\textit{N}& \cellcolor{purple}\textit{N} &\cellcolor{teal}\textit{N}  & \cellcolor{purple}\textit{N}\\
          
   \textbf{facebook-android-sdk} & 72 &   \cellcolor{teal}\textit{N} & \cellcolor{teal}\textit{N}& \cellcolor{teal}\textit{N} &\cellcolor{teal}\textit{N}  & \cellcolor{teal}\textit{N}\\
           
                     \textbf{lombok} & 50 &   \cellcolor{purple}\textit{N} & \cellcolor{purple}\textit{N}& \cellcolor{purple}\textit{N} &\cellcolor{purple}\textit{N}  & \cellcolor{teal}\textit{N}\\
                     \textbf{jna} & 45 &   \cellcolor{teal}\textit{S} & \cellcolor{teal}\textit{S}& \cellcolor{teal}\textit{S} &\cellcolor{teal}\textit{S}  & \cellcolor{teal}\textit{N}\\
                    \textbf{hector} & 45 &   \cellcolor{teal}\textit{S} & \cellcolor{teal}\textit{S}& \cellcolor{teal}\textit{S} &\cellcolor{teal}\textit{S}  & \cellcolor{teal}\textit{N}\\   
                        \textbf{okhttp} & 41 &   \cellcolor{teal}\textit{S} & \cellcolor{teal}\textit{S}& \cellcolor{teal}\textit{S} &\cellcolor{teal}\textit{S}  & \cellcolor{teal}\textit{S}\\

                    \textbf{commons-io} & 20 &   \cellcolor{teal}\textit{N} & \cellcolor{teal}\textit{N}& \cellcolor{teal}\textit{S} &\cellcolor{teal}\textit{N}  & \cellcolor{teal}\textit{N}\\

        \bottomrule
    \end{tabular}
    \end{center}
\end{table*}

We have used the JavaSymbolSolver~\cite{symbol} to produce the call graphs of our 49 projects. Producing the call graphs for every commits of a project is difficult and very time consuming. We will need to produce $\sim$25,000 call graphs for the hadoop project alone, because of its $\sim$25,000 commits. As a result, we selected only one commit per project for building the call graph and tracking the revision history of the project from that commit. Therefore, each project was checked out to two years before its current commit (current commit is presented in Table~\ref{tbl:subjectsTable}). This 2 years provides enough time to all the methods to undergo changes. 

The JavaSymbolSolver, unfortunately, was not able to produce call graphs for some of our projects including elasticsearch, flink etc., mostly due to \texttt{StackOverFlowError} exceptions. From the \emph{issue} tracking system, we found this problem common for the JavaSymbolSolver~\cite{symbolproblem}. Additionally, after removing entries with unmatched SLOCs, projects with very small number of methods, such as junit4, had too few samples to compare the maintenance indicators among the individual and the merged methods.  After filtering out these two categories of projects from analysis, Table~\ref{tb:colored-table} shows the results for the remaining projects. 
Most of the time the \textit{candidate for merge} methods were less expensive than the individual methods (i.e., more deep-blue cells than purple cells). More specifically, out of 115 cells, 84 of them are blue (73\%). If we consider cells with non-negligible effect size, the percent of blue cell becomes 87\%. This suggests that decomposing large methods will not increase maintenance effort of software systems. 
\begin{hassanbox}
\textbf{Summary:} Developers should decompose large methods so that none of the methods are larger than 24 SLOC.
A group of small methods, with sum SLOC \texttt{x}, are generally collectively less change- and bug-prone than an individual large method with SLOC \texttt{x}.   
\end{hassanbox}


\subsection{RQ4: Does controlling for size controls other quality metrics?}

So far, we have studied the impact of size on change- and bug-proneness. Change- and bug-proneness, however, are not sufficient for understanding software maintenance as a whole. In this paper, we consider three more code quality factors that could potentially impact the maintenance of a software project. 
In particular, we investigate, if controlling size (i.e., keeping method size within 24 SLOC) also controls these other three code quality metrics that have been associated with software maintenance. 

\textbf{1) Testability}: Developers and testers write test methods to evaluate whether a source method produces expected results. Writing test methods can be difficult for source methods with too many independent paths. Therefore, the popular McCabe metric is often used as an indication of the testability of a source method~\cite{Zhang:2013, Beer:2018}. 
The McCabe metric counts the number of independent paths in a code component with the formula: $1 + \#predicates$~\cite{McCabe:1976}. 
McCabe, however,  does not consider the number of control variables in a predicate. This is a limitation, because a predicate with more control variables should be considered more complex~\cite{Kafura:1987} and less testable than a predicate with a single control variable. 
McClure~\cite{McClure:1978}, on the other hand, considers all the control variables and comparisons in a code metric, and can be used as an alternative of McCabe. We consider both McCabe and McClure as indications of the testability of a source method. 

\begin{figure*}[htbp]
\centering
\includegraphics[width=1\textwidth ,keepaspectratio]{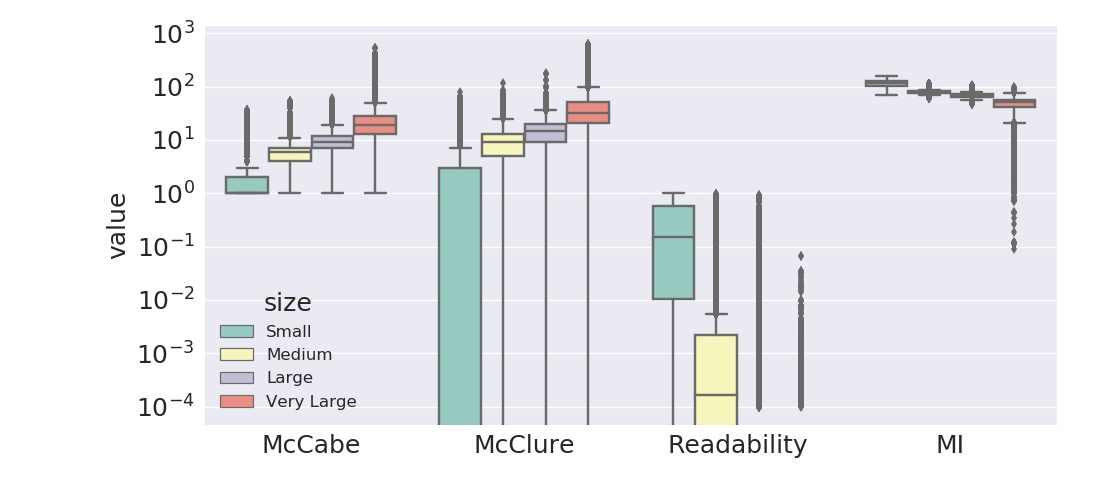}

 \vspace{2.5mm}
 \caption{For the Elasticsearch project, this graph compares the distributions of different quality metrics after grouping them according to the four different method sizes. A high value for McCabe (and McClure) indicates a high maintenance effort, whereas a high readability (and maintenance index) value indicates low maintenance effort. 
 } 
 \label{fig:quality}
 \end{figure*}

\begin{table*}[htbp]
\begin{center}
    \caption{Cliff's Delta Effect sizes after comparing methods in different size categories (e.g., Small-Medium indicates comparison between small and medium size methods). Effect sizes are N$\sim$Negligible, S$\sim$Small, M$\sim$Medium, and L$\sim$Large. For example, the first column of the first row compares the McCabe distributions between small and medium size methods and shows the percent of differences with different effect sizes.}
     \label{tbl:quality-effect-sizes}
    \begin{tabular}{ l c| c| c| c| c| c| c| c|c|c|c|cc }
        \toprule
        & \multicolumn{4}{c|}{\textbf{Small-Medium}} & \multicolumn{4}{c|}{\textbf{Medium-Large}} & \multicolumn{4}{c}{\textbf{Large-Very Large}}\\
        
        \textbf{Metrics} & \texttt{N}&\texttt{S}&\texttt{M}&\texttt{L} & \texttt{N}&\texttt{S}&\texttt{M}&\texttt{L} & \texttt{N}&\texttt{S}&\texttt{M}&\texttt{L} \\
        \midrule
        
\textbf{McCabe} &0.00&2.04&0.00&97.96&6.12&4.08&16.33&73.47&0.00&2.22&8.89&88.89 \\ 
\textbf{McClure} &0.00&2.04&0.00&97.96&2.04&4.08&59.18&34.69&0.00&8.89&17.78&73.33 \\ 
\textbf{Readability} &0.00&2.04&0.00&97.96&0.00&14.29&40.82&44.9&44.44&44.44&8.89&2.22 \\ 
\textbf{MI} &0.00&0.00&0.00&100.0&0.00&0.00&0.00&100.0&0.00&0.00&0.00&100.0 \\ 

        \bottomrule
    \end{tabular}
    \end{center}
\end{table*}

\textbf{2) Readability}: According to Martin Fowler, ``Any fool can write code that a computer can understand. Good programmers write code that humans can understand.'' One of the most frequent activities in software development and maintenance is \textit{code reading}~\cite{Scalabrino:2016}. Readability is a measure of how easy (or difficult) it is to read and understand a source code~\cite{Buse:2010, Posnett:2011, mchain:2016, Johnson:2019}, which has significant impact on code maintenance. In this paper, we have used the publicly available readability tool implemented by Buse \textit{et al.}~\cite{Buse:2010} to examine the relationship between method size and readability. 
This tool calculates a readability score for a given method from 0 (least readable) to 1 (completely readable code).

\textbf{3) Maintainability Index}:
We also include the maintainability index (MI) metric, which is industrially popular as a measurement of code quality. 
MI is calculated as: \\ \\
$171 - 5.2 * ln(Halstead\, Volume) - 0.23 * (Cyclomatic \, Complexity) - 16.2 * ln(Lines \, of \, Code) $ \\ \\ 
This is the evolved form of the original equation proposed by Oman and Hagemeister~\cite{Oman:1992}, and different forms of it have been adopted by popular tools such as Verifysoft technology~\cite{verifysoft}, and Visual Studio~\cite{studio}.

Considering the Elasticsearch project in Figure~\ref{fig:quality}, we can see that by controlling the method size, we can control the four above mentioned code quality factors in most cases. Similar to the RQ2, we compare these quality factors by grouping the method sizes. 
The Cliff's Delta effect size is large between all the compared method groups, except for the readability metric.
For the readability metric, the effect sizes are large, small, and negligible between \textbf{Small-Medium}, \textbf{Medium-Large}, and \textbf{Large-Very Large} method groups, respectively. Table~\ref{tbl:quality-effect-sizes} shows the findings for all the 49 projects. Similar to RQ2, the difference in the \textbf{Small-Medium} group is more meaningful than the others; the effect size for all of the four quality factors are large in at 97.96\% to 100\% of the projects. 
\begin{hassanbox}
\textbf{Summary:} By controlling size we can control other complex code quality metrics, such as McCabe, McClure, Readability, and Maintainability index.      
\end{hassanbox}

\section{Discussion}
Understanding code metrics and their relation with software maintenance is a forty years old problem~\cite{Lenarduzzi:2017}.
While there is debate about the effectiveness of many of the famous code metrics (e.g., C\&K), the research community is in complete agreement about the usefulness of size as an indicator of future maintenance. Therefore, research about size and maintenance is not new. It is obvious from all the previous studies that developers should keep the size of their code unit small, similar to our observation in RQ1. Unfortunately, there is no concrete evidence-based recommendations about how small the size should be. Although Visser \textit{et al.}~\cite{Visser:2016} recommended that developers should keep their method within 15 source lines of code, this recommendation was from intuition only, not based on evidence. Also, in RQ2, we have provided clear evidence that keeping method size always within 15 SLOC is less realistic.  

This paper is unique in that context, because we provide evidence-based recommendation that \emph{developers should keep their method size within 24 SLOC (RQ2)}. In RQ2, we also show that keeping method size $<$ 24 SLOC is less realistic, whereas method with size $>$ 24 SLOC is less maintainable. Therefore, this 24 SLOC bound can be used for trade-offs between feasibility and maintainability.  We also provide evidence that inherently large methods should be refactored to a group of helper methods, each within 24 SLOC (RQ3). Encouragingly, controlling for size, in most cases, 
controls other code quality factors, such as testability and readability (RQ4). 

Our findings can be used for improving code metric tools such as \textit{checkstyle}. For example, the current default limit of method size in \textit{checkstyle} is 150~\cite{checkstyle}. Our research shows that such a large method would be extremely maintenance prone, and this limit should be set to 24 only. Our methodology can be extended to find maintainable and feasible limit for other popular code metrics, such as McCabe, nested block depth, and maintainability index.

\section{Threats to validity}
Several threats, however, may impact the validity of our results. 

\emph{External validity} is hampered by our selection of software projects. 
Our observations with open-source software may not be consistent with closed-source projects.  
The selected 49 projects, however, are popular and have been widely used by other code metrics studies. 

\emph{Internal validity} is hampered by the statistical approaches. For example, the use of correlation coefficients might be inaccurate for unknown confounding factors. For finding the three critical values,  we have relied on the Alves \textit{et al.}'s approach. Other critical values may produce more useful results. 

\emph{Construct validity} is affected by our approach to define bug-proneness. Capturing bug related keywords from the commit messages for measuring bug-proneness is only $\sim$80\% accurate~\cite{Spadini:2018}. Also, the CodeShovel tool\cite{Grund:tool} we used for capturing method history is accurate only for 90\% of the methods~\cite{grund:2021}.  

\emph{Conclusion validity} is affected by the selected maintenance and quality indicators. Change-/bug-proneness, and the quality metrics we used may not be sufficient to understand software maintenance.
\section{Conclusion}
Significant research has been done to understand code metrics impact on software maintenance for building accurate predictive models. This paper investigated the relationship between method size and maintenance effort, because size has been frequently reported as the most common, easy-to-measure, and the most effective code metric for understanding maintenance. 
We established that developers should be careful about their method size for reducing maintenance effort, such as change- and bug-proneness. We also showed that controlling for method size, in most cases, controls other code quality metrics, such as testability, and readability. 
In general, developers should aim to keep their Java method within 24 source lines of code. They should also refactor inherently large methods to group of smaller methods, even if it increases the overall coupling of a system.  

The methodology of our study can be extended for other programming languages, and for other metrics (e.g., McCabe, and C\&K). Studies can also investigate other granularity, such as class/file and module levels. We hope such studies can help us better understand the relationship between code metrics and maintenance for building ever more accurate maintenance models. 

\section{Acknowledgments}
Shaiful Chowdhury was supported
by the Natural Sciences and Engineering Research Council
of Canada (NSERC, PDF-533056-2019). Gias Uddin was supported by an NSERC Discovery Grant (RGPIN-2021-02575).
\bibliographystyle{ACM-Reference-Format}
\bibliography{paper}
\end{document}